\documentclass[11pt]{amsart}
\usepackage[utf8]{inputenc}
 \usepackage{mathtools}
\def\dbar{{\mkern3mu\mathchar'26\mkern-12mu d}}
\usepackage{geometry}                
\usepackage{graphicx}
\usepackage{amssymb}
\usepackage{epstopdf}
\title{Maxwell's Demon}
                                     
\begin{document}
\maketitle

\centerline{R. E. Kastner}\smallskip
\centerline{\small {May 4, 2025; University of Maryland, College Park. rkastner@umd.edu}   }\medskip

ABSTRACT. This work provides an overview of key historical developments in the formulation of the Second Law of Thermodynamics,
focusing on the notorious challenge of ``Maxwell's Demon'', a hypothetical creature who could presumably violate that law. It begins by
recalling Maxwell's challenge and discussing the apparent loophole in the Second Law that appears to make such a violation possible.
An alternative formulation of the Demon challenge by Szilard is considered, along with his attempted defeat of the Demon through 
reference to measurement. A similar effort by Brillouin is also analyzed. The more recent proposal of Bennett to defeat the Demon
through the requirement of memory erasure is critically discussed. Finally, it is proposed that the Second Law gains a firm foundation through
 neglected features of quantum theory. In particular, an application of the Heisenberg Uncertainty Principle is shown to decisively defeat the Demon, as
 well as to serve as justification for Landauer's Principle, albeit in terms distinct from the usual computational formulation.

\section{Historical Background}

  What is ``Maxwell's Demon''? It is a thought experiment proposed by  James Clerk Maxwell  in 1867 upon observing what seemed to a loophole in the Second Law of Thermodynamics. The Second Law predicts that an isolated system will naturally evolve over time to an equilibrium state, such as a uniform temperature throughout a volume of gas. It is usually stated in terms of the thermodynamic quantity ``entropy'', which is a measure of heat $\Delta Q$ transferred at a particular temperature T: $\Delta S = \frac{\Delta Q}{T}$.  This definition of entropy in terms of heat is due to Clausius and has been empirically corroborated (Clausius 1867).
   An equilibrium state is a state of maximal entropy for a given system. Thus, any change in the system's entropy $\Delta S$ is always positive (or, if it is already at equilibrium, 
   $\Delta S=0$). The typical example of evolution towards maximal entropy is that of a volume of gas initially partitioned into two smaller volumes at different temperatures, with removal of the partition and subsequent evolution towards a uniform temperature throughout. Maxwell's Demon is a hypothetical creature who can do the reverse: start with a volume of gas at uniform temperature and effect microscopic changes resulting in two different temperatures on each side of the container. In Maxwell's words:

\begin{quote}

“Now let A and B be two vessels divided by a diaphragm and let them contain elastic molecules in a state of agitation which strike each other and the sides…. Now conceive a finite being who knows the paths and velocities of all the molecules by simple inspection but who can do no work except open and close a hole in the diaphragm by means of a slide without mass. Let him first observe the molecules in A and when he sees one coming the square of whose vel.[ocity] is less than the mean sq. [square] velocity of the molecules in B let him open the hole and let it go into B. Next let him watch for a molecule of B, the square of whose velocity is greater than the mean sq.[square] vel. [velocity] in A, and when it comes to the hole let him draw the slide and let it go into A, keeping the slide shut for all other molecules. Then the number of molecules in A and B are the same as at first, but the energy in A is increased and that in B diminished, that is, the hot system has got hotter and the cold colder and yet no work has been done, only the intelligence of a very observant and neat-fingered being has been employed. Or, in short, if the heat is the motion of finite portions of matter and if we can apply tools to such portions of matter so as to deal with them separately, then we can take advantage of the different motion of different proportions to restore a uniform hot system to unequal temperatures or to motions of large masses. Only we can’t, not being clever enough.” (Knott 1911, pp. 213–214, letter to Peter Guthrie Tait ) 

\end{quote}

What Maxwell describes above is the violation of the Second Law by a ``very observant and neat-fingered being.'' Thus, his ``being'', which came to be known as the Demon, represents an attempt at a counterexample to the Second Law; i.e., a demonstration that it should not be considered an exact, physically mandated result but is only approximately true and violable based on sufficient knowledge of the normally hidden microscopic properties of a physical system.

 We should note, since it will be highly relevant later on, the significance of Maxwell's characterization of the system's molecules as being ``in a state of agitation,'' since that is already ``smuggling in'' the notion of randomness due to what is often termed ``thermalization.''  The latter term roughly means ``reaching equilibrium through heat exchange.'' The concept of {\it heat} plays a central role in the study of entropy and its changes (as is obvious from the definition above). Heat is often defined in such terms as ``thermal energy transferred between systems at different temperatures.'' If one inquires as to what is meant by ``thermal energy'', one arrives at definitions involving ``energy of random (or disordered) motion'' and observations that such energy is not available for ``useful work''.  However, against the backdrop of classical mechanics, what counts as ``disordered'' vs. ``useful'' is apparently scale-dependent, and thus observer-dependent. Thus, before attending to further details, we wish to flag the historically ill-defined nature of these basic concepts. Their lack of an observer-independent definition (at least under deterministic classical mechanics) is an aspect of the apparent loophole in the Second Law available for exploitation by a ``Demon"'. 

\subsection{Fundamental Questions}

 In what follows, we will provide some historical background on the discovery and formulation of the Second Law, including some crucial background assumptions that have served to make the Demon a conceivable and significant challenge. We then address the fundamental questions proposed by Hemmo and Shenker (2016):
 \begin{quote}
 ...There are two major questions about Maxwell’s Demon: (I) Is the Demon compatible with the principles of fundamental physics? (II) Are there, or can one construct, Demons in our world?  
 \end{quote}
 
Hemmo and Shenker assert that the answers to both questions is ``yes"; i.e., they argue that such a being is possible and could defeat the Second Law in a way compatible with the laws of physics. The present review will show that this kind of affirmative response arises from an epistemic reading of the statistical treatment of physical systems--i.e., the assumption that probabilistic uncertainty does not describe an uncertainty in system properties but only the ignorance of a typical observer, which opens the way to an {\it atypical} observer (the ``Demon'') who is not ignorant. However we will also discuss the possibility of a negative response---i.e., that no amount of intelligence and neat-fingered-ness can defeat the Second Law. In what follows, we will show that a rigorous physical refutation of the Demon is available based on an ontological reading of statistical uncertainty, i.e., an uncertainty in the properties (such as position or energy) possessed by the systems themselves, along with real physical limitations on the degree to which those properties can be manipulated via the use of measurement. Of course, both the ontological reading of statistical uncertainty and any limitation on measurement can only come from the quantum level, since classical physics takes all system properties as intrinsically determinate and treats ascertainment of those properties as trivial by inspection that does not affect the systems. Thus, in Maxwell's time, the only theoretically supported reading of the statistical postulate was an epistemic one, which seems to make the Demon hard to refute. The view of Hemmo and Shenker that a Demon is possible, as well as longstanding and often-repeated claims that the only way the defeat the Demon is through erasure of its memory (where it is assumed to have unrestricted access to properties of systems; e.g., Bennett, 1982; 2003) shows that the epistemic (ignorance-based) interpretation of probabilities appearing in statistical thermodynamics remains endemic in the debate over the Demon to the present day. In contrast to this conventional view, a major aim of the present overview is to make clear the availability of a rigorous, quantum-level defeat of the Demon that has evidently not been taken into account in extant analyses. The latter is based on an understanding of the probabilities as ontic---i.e., as reflecting objective physical uncertainty concerning a system's micro-properties.

Before proceeding further, some historical context concerning the Second Law is relevant and helpful here. First, some terminology: the relevant states of systems, in this case a gas, are called ``macrostates''; these are characterized by parameters such as number of molecules N, volume of the system V, temperature T, and total energy of the system E. In contrast, a``microstate'' is a detailed description at the level of the individual components, such as ``how many molecules have velocity $v$ at time $t$.''

\subsection{Stochasticity}

Since gas molecules are always colliding, these velocities change rapidly and the relevant quantity becomes ``the probability at some time $t$ that $n$ molecules have velocity $v$.''  In the time frame of Maxwell's ``Demon'' proposal, the Second Law was in the form of a finding that a system will naturally evolve toward a stationary-over-time macrostate, which is called the ``equilibrium'' state. In particular, Maxwell had recently derived what is now called the ``Maxwell distribution" of velocities as the equilibrium state of a gas at temperature T. In terms of kinetic energy $E=\frac{1}{2}mv^2$, it takes the form:

$$ P(E) = Ae^{-\frac{E}{kT}} \eqno(1)$$

\noindent where $k$ is Boltzmann's constant and A is a normalization constant such that integration of the probability distribution over all velocities  yields unity.

 Maxwell's proposal of his ``Demon" challenge was in response to the idea that the Second Law seemed ``only statistical''; i.e., not a strict mechanical result
 following from the precise details of molecular motion, which in classical physics is deterministic and time-reversible. Indeed, Maxwell's equilibrium state (1) was arrived at by an assumption of randomness concerning molecular motion, which treats transitions among the gas microstates as Markov processes--i.e., processes in which a system's state jumps randomly from one state to another according to a stochastic (indeterministic) law. At the classical level, since the microscopic laws are fully deterministic, the only way to get  apparent stochasticity is by not taking into account the full details; i.e., by appealing to the  ignorance of the typical observer of the full, allegedly deterministic, details. This is also often referred to as ``coarse-graining'': an averaging over finer-level details that are taken as determinate but unknown or irrelevant. It is this ignorance assumption that opens the loophole for a ``Maxwell's Demon'': i.e., an observer who is not ignorant of the full details and who, at least conceivably, has the capacity to intervene in what would otherwise apparently be an inexorable evolution towards equilibrium. 
  
 The randomness assumption was later adopted by Ludwig Boltzmann [1872], who called it the ``Sto\ss zahlansatz'' or SZA for short. He used the SZA in his ``H-Theorem'', which was a proof, based on certain assumptions including the SZA,  that entropy increases for any closed system not in an equilibrium state and thus a proof of the Second Law. The SZA is crucial for this derivation, which has been subject to legitimate criticism from the standpoint of classical physics (e.g. the Loschmidt reversibility objection, Loschmidt 1876, 139). In brief, Loschmidt's objection is that for any evolution of the system in which entropy increases, there is an equally valid one obtained through reversal of all velocities, in which the entropy decreases. Since under classical mechanics the dynamics is assumed to be reversible, one cannot rule this out. Thus, under classical micro-dynamics there is no reason for entropy increase to outweigh entropy decrease. In addition, the physical meaning of the SZA was unclear in ensuing developments attempting to place the Second Law on more rigorous footing. In particular, there was equivocation as to whether the probabilities were meant to describe temporal uncertainty of a system's state or uncertainty concerning occupation number N of gas molecules of the various energy states. (For a careful historical analysis of this equivocation, cf. Uffink, 2024). For additional historical background on the topic of Maxwell's Demon, cf. Leff and Rex (2003).
 
  It is impossible to overstate the significance of the SZA for the ensuing massive amount of debate over the status of the Second Law and its challenge by Maxwell's Demon.  Despite the tenuousness of its basis in the context of classical physics, the SZA in the form of the assumption of a Markovian process is the ``engine'' that gets the Second Law ``off the ground.'' It is argued in Kastner (2017) that rigorous justification can be found for the SZA only at the quantum level, which provides a basis for ontological uncertainty concerning a system's properties and a discontinuous ``quantum jump'' applying to transitions among microstates. 
  
   In short, we will find that if the SZA strictly holds at a fundamental physical level, rather than being based only on ignorance of a typical observer, the probabilities become ontological in nature and the Second Law gains a physically objective grounding. This alone does not preclude a Demon from intervening in the process towards equilibrium; but we also find that taking into account the quantum uncertainty principle vindicates the proposals of Szilard (1929) and Brillouin (1951) that it is the process of measurement that saves the Second Law from the Demon. (This is not to say, however, that either Szilard or Brillouin made a good case for measurement as the savior of the Second Law; Szilard's proposal was circular, and Brillouin neglected to take into account the quantum uncertainty principle even as he pointed to an energy cost for measurement.)

In order to understand the crucial role of the SZA for support of the Second Law, and the implications for the viability of Maxwell's Demon, we now turn to the specifics of Boltzmann's H-theorem.

\section{Boltzmann's H-theorem}

The Second Law was already present theoretically in nascent form through Maxwell's velocity distribution (1). However, it acquired a more general footing through Boltzmann's ``H-theorem,'' (Boltzmann, 1872). This theorem proposed a quantity, the ``H-function'' that
tended to a minimum in a closed system subject to stochastic transitions among microstates. The H-function is essentially the negative of entropy ``S'' as defined in terms of probabilities $p_i$ of microstates,\{$i$\}, i.e.: 

$$S =   -k\  \sum_{i} \ p_i  \ln \ p_i \eqno(2) $$

\noindent where k is Boltzmann's constant.  (Boltzmann's original form takes a continuous phase space distribution and integrates, but we use a discretized version for applicability to the quantum form.) The equivalence of (2) to Clausius' definition of entropy is found through the relation $\Omega = e^{\frac{S}{k}}$
where $\Omega$ is the number of microstates consistent with the macrostate having entropy $S$.  In terms of specific microstates with varying probabilities, the more general formula for S is (2).  Boltzmann arrived at the form of the H-function by noting that it is minimized for the Maxwell distribution. While the Clausius entropy applies only to equilibrium situations, the expression (2) is more general and applies to non-equilibrium distributions as well.

One investigates how S will change with time by differentiating the above and finding:

$$ \frac {dS}{dt} = -k\  \sum_{i}  (  \frac{dp_i}{dt}  \ln p_i + \frac{dp_i}{dt} ) = -k\  \sum_{i} \frac{dp_i}{dt} \ln p_i \eqno(3)$$

\noindent where the second term is zero since $\sum_{i} p_i =1$. Now, to evaluate $\frac{dp_i}{dt}$,
we need transition rates $r_{ij}$ from state $i$ to state $j$ and vice versa.  In the classical form of Boltzmann's theorem, one employs epistemic
ignorance (``coarse-graining'') over the assumed deterministic trajectories to obtain these rates, and finds that $\frac{dS}{dt}$ is always either
positive or zero. Since the same basic calculation applies at the quantum level, we do the explicit evaluation in that context.

\subsection{The H-Theorem in Quantum Form}

Boltzmann's derivation can be cast in a form suitable for application to quantum mechanics if the transition rates are those obtained from Fermi's Golden Rule (the latter being an application of the Born Rule under a perturbation causing transitions between particular initial and final states). Since Fermi's Golden Rule is naturally symmetrical under reversal of the initial and final states (as is the Born Rule), one finds that $r_{ij} = r_{ji}$.
 The result is often called a ``master equation'' and takes the form:

$$ \frac{dp_i}{dt} = \sum_{j} r_{ij} p_j - r_{ji} p_i  \eqno(4)  $$

  In matrix form, this looks like:

$$ \frac{d\bf{p}}{dt} = -[R] \bf{p} \eqno(5) $$

\noindent  where $[R]$  is the ``master operator''.  Its elements are given by $R_{ij} = r_{ij}$ for $j \ne i$ and $R_{ii} = -\sum_{j\ne i} R_{ij}$.

The solution in matrix form is

$$ \bf{p}(t) = exp^{-[R]t }\bf{p}(0) + \bf{p}^{eq}  \eqno(6) $$,

\noindent which gives rise to a decaying time dependence leading to equilibrium. The equilibrium distribution $\bf{p}^{eq}$ is that for which 
$\frac{d\bf{p}^{eq}}{dt} = 0$, and this arises from the detailed balance condition which requires  $r_{ij} p_j^{eq} = r_{ji} p_i^{eq}$.
Plugging the latter condition into (4) confirms this point.

This analysis already shows that a system described by this sort of stochastic mechanics will inexorably evolve to
an equilibrium state, implying maximal entropy, but we now show that explicitly. Substituting the expression (4) into  (3), we find

$$ \frac {dS}{dt} = \frac{k}{2}\  \sum_{i,j} r_{ij} ( \ln p_j - \ln p_i ) (p_j - p_i) \eqno(7) $$

\noindent We now observe that the two differences will always be of the same sign, so that $ \frac {dS}{dt} \ge 0$.

\subsection{Criticisms and Implications}

As noted above, Boltzmann's derivation assumed ``molecular chaos'' (the SZA) that could not be grounded in
classical physical laws, and that led to a plethora of criticisms that were fully legitimate against the backdrop of classical physics.
 The most well-known is that of Loschmidt, who (as mentioned above) pointed out that under classical mechanics, any micro-evolution of the system satisfying $ \frac{dS}{dt} > 0$ can be reversed to an equally probable one leading to $\frac{dS}{dt} < 0$. Burbury (1894) probably deserves the most credit for
pinpointing the actual source of the irreversibility under the H-theorem, and arguably it was not a specific assumption of time-asymmetry but rather
an assumption of true randomness that could not be supported under classical laws. Burbury called this ``condition A'' and
he put it as follows concerning a collision of two particles: ``For any given direction [of mutual velocity] before collision, all
directions after collision are equally probable.'' (Burbury 1894, 78).  He then noted that reversing the velocities and attempting
to show that the H-theorem fails actually amounts to applying it in a way that violates its basic assumption ``A''; namely, a randomness
in the evolution of physical systems. Thus, randomness, or in more precise terms, a Markovian (indeterministic, stochastic) process, is the crucial assumption
of the H-theorem. (This stochasticity generalizes to situations beyond the equiprobable situation mentioned by Burbury.)  But that of course denies the determinism of classical mechanics, which decrees that the evolution of a system (in either temporal direction) is uniquely dictated by its current state and the applicable deterministic law.
 It is thus denial of determinism of the evolution of the particles that is the crucial assumption that gives the H-theorem its power to yield a temporally directed entropy increase. That result is being extracted from stochasticity, not explicit time-asymmetry.
 
  Burbury goes on to acknowledge the fact that this seems to conflict with classical mechanics, but comments:
 
 \begin{quote}
 
 Any actual material system receives disturbances from without, the effect of which, coming at haphazard, is to produce
 the very distribution of coordinates which is required to make [S increase]. (Burbury 1894, 78)
 
 \end{quote}
 
 This is an interesting statement in that it asserts that the ``actual'' world is haphazard, despite classical mechanics.
 Since the finding of the H-theorem is consistent with empirical phenomena, physicists of the time, like Burbury, had to
 view it as capturing something about the world despite its basic assumption  --the SZA -- that conflicted with a basic
principle (determinism) of classical physics. 
As Uffink notes: ``One thing is certain, and that is that any such preference [for temporally directed irreversibility] cannot be obtained from [classical] mechanics and probability theory alone.''  (Uffink, 2024) What it can come from is a physically real Markov process,
since that is the essential element of the SZA (or Burbury's ``condition A''). Comprehensive studies of the controversy around the H-theorem
 can be found in Uffink (2001, 2007, 2024) and
Uffink and Valente (2010, 2015).\footnote{In their (2015), Uffink and Valente argue that there is actually no time-asymmetric ingredient in Lanford's Theorem (1976) either,
which agrees with our conclusion here that the essence of the SZA is statistical independence arising from a fundamental stochastic dynamics.}

Thus, the H-theorem is corroborated by empirical phenomena, yet its key assumption violates fundamental features of classical mechanics (determinism and reversibility). It is not usually put in these terms, but the situation is arguably analogous to 
Planck's derivation of the blackbody radiation law from an assumption (discreteness of energy) that violates the classical theory. While Planck
originally viewed $E=h\nu$ as a calculational device, it was eventually acknowledged (with the help of Einstein's analysis that we address
in further detail below) that the classical theory is inapplicable in specific ways and requires correction at the quantum level. In parallel, one
might reasonably expect that the empirical success of the H-theorem should lead us to consider that perhaps the deterministic and time-reversible features of classical mechanics require updating and correction (as is suggested by Burbury's observation that ``actual mechanical systems'' 
are subject to stochasticity). Yet in most discussions of Boltzmann's H-theorem, it is still generally assumed that the evolution of systems ``really is'' classically deterministic (at least as a good approximation) and that the SZA is a weakness of the proof. The present review encourages a reconsideration
of this usual picture, proposing that the SZA is not an illicit smuggling-in
of irreversibility but in fact reflects the ontology of the world, just as Planck's ``trick'' $E=h\nu$ reflects the ontology of the world (i,e., that photons
exist and their energy is in fact $E=h\nu$ ). Another way to put this is that Burbury is correct in the quote above in saying that ``actual systems" are subject to ``haphazard'' influences, even if he perhaps did not realize (or might not agree) that he was suggesting that the world is not really classically deterministic. 

We turn next to the relevance of Markov-type processes, involving ``heat,'' the form of energy that does the ``heavy lifting'' in defining
entropy and its evolution. But first, we entertain a brief aside examining the concept of heat as traditionally invoked in a thermodynamic context.

\subsection{Heat vs. Work: an ambiguous distinction?}

 Both heat $\dbar Q$ and work $\dbar W$ are inexact differentials, in that they are path-dependent. And this is no coincidence, because
 if we look carefully at the relevant details, we find that, arguably, there is no fundamental difference between the two quantities.
  ``Heat'' is often defined as energy that is only transferred from a hotter system to a colder one, and
 ``work'' is used in connection with an applied force acting over a distance that could transfer energy from a colder to a hotter one; thus, at first glance, they appear to be distinct quantities.  Strictly speaking, a finite quantity of heat, $\Delta Q$, is only defined in terms of a corresponding increase
 in temperature:  $\Delta Q = m C \Delta T $, where m is the mass and $C$ the system's heat capacity.  However, T is defined in terms
 of the average kinetic energy of molecules, and molecules can only have their kinetic energy changed by doing work on other
 systems or having other systems do work on them. Thus, the very definitions of these quantities are inseparable form one another. 
 
  Moreover, for a system in equilibrium, one continues to have ``thermal interactions''; that is, ``disordered
 motion of molecules'' in which energy is exchanged among the molecules. That energy is neither being transferred ``from a 
 hotter system to a colder one'' nor is it a quantity of {\it macroscopic} work such as $p dV$. It is then characterized as the internal energy of the system,
 $E = \frac{3}{2}NkT$.  However, since the notion of ``uncontrolled'' or ``disordered'' motion is often associated with heat, it's important to
 note that under classical mechanics, no motion is truly ``disordered,'' since it is supposed to follow deterministic laws. Besides the 
 apparent inconsistency of the nature of heat with these laws, the notion of controllability brings in intentional meanings; i.e., that work must be defined in terms of intentionality and agency while heat, in contrast, is typically defined in terms of absence of agency. The intentionality extends
 to the definition of what counts as the ``system'' whose temperature is relevant.  Yet, even in an equilibrium state, one cannot deny that a particular molecule A in some excited state is emitting a photon to another molecule B, which (for typical energies at T) will thereby be accelerated, and that therefore molecule A is doing a quantity of work on molecule B, $\Delta W = \frac{1}{2}m (v_f^2-v_0^2)$, even if that work
 does not result in a ``system''-level change. The point of this example is that definitions of these alternative ``forms'' of energy are observer-dependent
 and that there is no absolute distinction between ``heat'' and ``work''. From the point of view of a human, the interaction of the molecules is
 ``uncontrolled' and is called ``heat''; from the point of view of the molecules, one molecule is quantifiably doing work on the other to accelerate it, whether or not molecule A had any intention of doing so, or whether it was a ``controlled'' act.
 
Thus, in ``thermal'' interactions at equilibrium, molecules are still doing work on one another. We can call it ``micro-work'' if we wish, to indicate that it is not resulting in changes in macroscopic parameters like volume, pressure and temperature. But energy is being transferred at the molecular
level.\footnote{Of course, this reminds us that what counts as the ``system'' whose volume is relevant remains scale-dependent and thus observer-dependent: suppose the molecule A is a microscopic scientist and he is working with a single molecule B that is his system. His injection of energy into the system of molecule B increases the volume of that system!}  This leads us to consider in more detail the conditions for an apparent ``uncontrollability'' of energy exchange that we call ``heat'' and its relation to the indeterminism that is actually required for a firm foundation for the Second Law. 
 
\section{The Quantum level as a Source of Stochasticity} 

\subsection{``Those Damned Quantum Jumps"}

The quote comes from a famous comment from Erwin Schrodinger to Niels Bohr.  Schrodinger was decrying the empirically observed discontinuities associated with transitions among energy levels in atoms, since these were evidently inconsistent with the (unitary) deterministic evolution
of his eponymous equation. The epigraph opens an essay by John S. Bell (1987), who argued that there is good reason
to suppose that the jumps are real and not something to be avoided but which, even according to an earlier comment
by Schrodinger himself, could well be a fundamental aspect of reality that is not captured by the wave equation. Bell refers in particular to this 1922 quote, recalled later in Schrodinger (1957):

\begin{quote}
Once we have discarded our rooted predilection for absolute Causality, we shall succeed in overcoming the difficulties.
(Schrodinger, 1929)
\end{quote}

Schrodinger's evident ambivalence about the issue, along with Bell's observation in his (1987) that perhaps the wave equation
does not capture everything needed for a full quantum theory and that there is indeed physical collapse or reduction, provides
some historical background for the idea that genuine stochasticity is required not just for consistency with empirical phenomena but also for a consistent physical grounding
of the Second Law.  Let us now consider two important cases of empirical evidence for ``damned quantum jumps''.

\subsection{Brownian Motion}

 In this section, we consider a well-known empirical phenomenon that is inconsistent with classical mechanics, and forced
 researchers to deal with apparently genuine physical indeterminism in Nature that stems from the quantum level,
 including ``damned quantum jumps.'' As is well known, botanist Robert Brown discovered the phenomenon in 1827 while
 examining pollen grains in water under high magnification. What he saw was a ``jittery motion'' that he initially thought
 was connected to the biological (living) aspect of the system, but upon repeating the experiment with inanimate matter
 (e.g. dust particles) he found the process still occurring. As is also well known, Einstein (1905a) provided an analysis
 that succeeded in accounting for the observed motions in terms of an underlying atomic structure (water molecules) undergoing stochastic
 dynamics. Specifically, Einstein showed that the larger pollen grains could be modeled as undergoing a random walk (Markov process) 
 due to collisions with atoms, whose motions are taken as stochastic. The dynamics is that of a diffusion equation. The importance of stochastic fluctuations in this process is indicated by the fact that the pollen grains (and  the water molecules) undergo a net displacement over time, despite the
 fact that individual collisions in either direction are equally probable. This displacement arises from the second moment of the
 probability distribution, which can only be non-vanishing if there are random fluctuations in the motions of the molecules.
 
 Of course, in discussions of this and related diffusion phenomena, it is typical to portray the randomness as corresponding to ignorance of tacitly assumed underlying classical trajectories. However, it would seem that if such were the case, a model based on deterministic dynamics
 would be possible in principle, and could be constructed for a number of molecules N of finite size, even if N is large.  However, given
 the reversibility of the assumed dynamics, and the fact that diffusion is an irreversible process, this is a logical impossibility facing the
 same objections as the H-theorem under the assumption of micro-reversibility. Thus, all viable models of diffusion phenomena such as Brownian motion  require the stochasticity assumption, suggesting that it is a physical property of the systems. We do, of course, have ``classical diffusion equations,'' such as Fick's Law (Fick 1855) but these are always based on an initial probability distribution, so that stochasticity is ``smuggled in'' at the outset, and moreover is a continuum approximation that fails at the quantum level. At that point, ``random walk'' models are employed which can take into account molecular-level details. The logical impossibility of deriving an irreversible process form reversible dynamics, along with the plethora of empirical phenomena that obey stochastic models, reinforces the idea that stochasticity (i.e. indeterminism) is an ineliminable property of the physics. 

\subsection{Spontaneous Emission and Fermi's Golden Rule}

A specific  quantum-level process that manifests stochasticity, and attendant irreversibility, is that of spontaneous emission. Spontaneous emission is a non-classical process
in which an excited atom or molecule transitions from its excited state into a lower energy state in the absence of stimulation by an external coherent field, the
latter being the classical form of the field. Instead, owing to to quantum nature of bosons (in this case, photons), there is non-vanishing amplitude for a real 
(on-shell) photon to be created, despite the absence of other real photons in the field. Einstein (1916) was the first to predict this effect, as recounted in 
Straumann (2017), who notes: ``[Einstein] shows that in every elementary process of radiation, and in particular in spontaneous emission, an amount $\frac{h\nu}{c}$ of momentum is emitted in a random direction and that the atomic system suffers a corresponding recoil in the opposite direction.'' The key point here is that the process is ``random,'' or indeterministic, which  Einstein somewhat ruefully notes: 

\begin{quote}

The weakness of the theory lies, on the one hand, in the fact that it
does not bring us any closer to a merger with the undulatory theory,
and, on the other hand, in the fact that it leaves the time and direction of elementary processes to ‘chance’; in spite of this I harbor full
confidence in the trustworthiness of the path entered upon.

\end{quote}

Of course, Einstein's aversion to a ``dice-playing God'' is well known, and that is the basis for his characterization of this indeterminism as a ``weakness''.
 However, he also admits that the ``path'' is the correct one and indeed the effect was later experimentally confirmed by Frisch (1933). 
 
Fermi's Golden Rule (FGR) is a result from quantum theory that allows prediction of decay rates, via spontaneous emission, of systems in excited states.
It is obtained from evaluating the transition amplitude between initial and final energy levels resulting from a first-order perturbation due to the relevant field, and then applying
the Born Rule to that amplitude to obtain a time-dependent probability $P(t)$.  In the limit of times long compared to the lifetime of the 
excited state, $P(t)$ approaches a constant, which is interpreted as a detection rate $\Gamma(E_i, E_f)$:

$$ \Gamma(E_i, E_f)  \equiv  \frac{P(t)_{i \rightarrow f}}{t} = \frac{2\pi}{\bar{h}}  |\langle i |H_I | f \rangle |^2  \delta(E_i - E_f - E_\gamma) \eqno(8)$$

In (8),  $H_I$ is the spatial part of the perturbing potential, and the delta function reflects energy conservation via the radiation of a real (on-shell) photon of 
$E_\gamma = h \nu$. FGR is well corroborated and is a mainstay of experimental investigations of excited states. For decays of excited atomic or molecular states through
photon radiation, the relevant perturbation is due to the electromagnetic field.
 
 An important observation is that the perturbation affects the value of the excited energy level in two distinct ways: (1) it changes the value of the energy level (shift in $E$)
 and (2) it introduces a spread $i\Delta E$ in the shifted energy level. (1) is exemplified by the Lamb shift (Lamb and Retherford, 1947) while (2)  characterizes the
 amplitude for decays and is the basis for Fermi's Golden Rule.  One obtains an overall decay rate $\Gamma$ of a particular excited state by summing over all allowed transitions between the initial excited state and final states. Specifically, $\Gamma = \frac{1}{\tau}$ where $\tau \sim \frac{h}{\Delta E}$ is the average lifetime of the state.
  Thus, decays and accompanying radiated quanta, such as real (on-shell) photons, are not possible without a finite width $\Delta E$ of the decaying energy level. As noted above, this width turns out to be an imaginary quantity, in contrast to the shift in the energy level that results from the same perturbation.
While most textbook derivations of FGR treat the imaginary nature of the line width as a calculational artifact rather than a physical property of the energy level, in fact
that aspect of the energy level signals a departure from unitarity since it renders the associated action complex.

We can see that the action governing perturbative changes in energy levels (including decay widths) is complex by expressing the quantum electromagnetic field in terms of the Feynman propagator:

$$ A_\mu (x) = \int d^4x\   D_F(x-y) J_\mu (y) \eqno (9) $$

\noindent where $D_F(x)$ has the form $D_F = \frac{1}{(2\pi)^4} \int d^4k \ (\frac{(PP)}{k^2} - i\pi \delta(k^2)) $, ``PP" being the principal part. 
Thus $D_F$ is complex, where the real (principal) part corresponds to virtual (``off-shell'') photons (applicable to internal scattering processes) and the imaginary part, the delta
function term, corresponds to real (``on-shell'') photons. It is only the latter which are involved in radiative processes. Thus, from this decomposition
of $D_F$ we see that real photons are associated with the imaginary component.

Inserting ($D_F$) into the action $S$ describing interacting currents yields:

$$S = \int d^4x \ J^{\mu}(x) A_\mu(x) = \int d^4x \ d^4y J_\mu(x) D_F(x-y) J^\mu(y) \sim
 \int d^4x \ d^4y \  d^4k\  J_\mu(x)  (\ \frac{PP}{k^2} - i\pi \delta(k^2) ) J^\mu(y) 
 \eqno (10)$$ 

We see from (10) that the second, imaginary term renders the action complex, and in fact this is the term corresponding to decays.
Complexity of the action implies physical non-unitarity (and associated indeterminism of the evolution) if the radiated photon is absorbed by a subset of all currents, which is what
we find empirically--i.e., given an excited source for which many absorbing currents are available, only one current receives the photon.
Indeed, this is the essence of the light quantum and the key feature of Einstein's prediction of the photoelectric effect (Einstein, 1905b).\footnote{The issue of how irreversibility is related to stochasticity and its various forms (including non-Markov stochasticity) is a subtle one. Different forms of decoherence hold under different circumstances, such as the relation of the lifetimes of the system's states vs. environmental system lifetimes, as well as temperatures. The basic point is that genuinely irreversible dissipative effects arise only through complex actions; i.e., through processes that involve the creation of an imaginary component of the relevant energy levels.}

\subsection{Mielnik's critique}

B. Mielnik has pointed out that the conventional treatment of quantum theory as involving
only unitary evolution remains afflicted with the problem of measurement and related inconsistencies and paradoxes. In his (2017),
he reminds us that these problems have not gone away for the form of the theory that that views
the dynamics as truly reversible at the micro-level, and notes their relevance to the status of the Second Law:

\begin{quote}

One of widely announced answers [to the Schrodinger Cat Paradox] is that the mechanism of wave packet reductions, “these
damned quantum jumps”, during the measurements are indeed the decoherence effects of the
dynamical evolution, leading always to an entropy increase. In fact, some increase of disorder
in our prediction of the future states is rather obvious. The data about the present state of either
classical or quantum particle system, even if good, are never infinitely exact. So even the little
errors of our initial knowledge might cause an increasing blurring in our predictions of the
future. The problem, however, is whether this lost of information is in any way responsible for
the sudden state jumps (interpretable rather as an escape from the disorder?)
By trying to understand the phenomenon in an arbitrarily complex but still micro-reversible
theory, one inevitably faces the seldom remembered Loschmidt paradox. It states that while the
prediction of the future might be increasingly blurred (entropy increase!) then if the theory
is micro-reversible, i.e. the change of $t$ into $-t$ is irrelevant for the dynamical laws, the disorder
(blurring) has the tendency to grow either if we try to predict the future or to reconstruct
the past....

The paradox makes senseless the persistent belief of growing disorder (entropy) in a large
class of physical theories. It includes all known cases of classical (multi-particle) dynamics
without friction. No different in quantum theories. No matter, how many subsystems describes
a quantum state, in all customarily considered cases of quantum dynamics, the evolution `toward
the past’, after a simple CPT conjugation, is identical to the `motion to the future.' ..
 
Hence, if the unitary evolution is the only element of the story, there is no such miracle [i.e. reduction
of the wave packet], that one of these directions will introduce more disorder than the other....
The situation can be different if the unitary evolution is not the only law, but is completed by some
additional mechanism. Only then, it might explain the growing decoherence in the future. Yet,
the question still remains, whether apart of the growing disorder (entropy) it will indeed describe
the sudden ‘reduction jumps’ of quantum states? (B. Mielnik, 2017, 2-3)

\end{quote}

 Mielnik goes on to mention that Lindblad-type dissipation can account for an arrow of time and entropy increase,
 remarking: ``the Lindblad dissipative term... privileges
the direction of time....Would it represent the missing element of the theory?''

What Mielnik is pointing out is that there can be no real discontinuous ``jump'' corresponding to the kinds of empirical
phenomena we observe (e.g. in Brownian motion and in detection rates as predicted by Fermi's Golden Rule), nor any corresponding entropy increase, without some genuinely dissipative process in quantum theory. Lindblad equations describe Markov processes; thus Mielnik quite appropriately
suggests that such processes are the currently missing elements of a quantum theory that can remedy the ``senselessness'' (his term)
of a belief in entropy increase against the backdrop of a micro-reversible theory. In short, what is needed is micro-irreversibility
corresponding to ``those damned quantum jumps.''

\subsection{Einstein's proof that photons behave as quanta undergoing random motion.}

Einstein's landmark paper on the quantum nature of light (1905b) did more than predict the now-famous photoelectric effect. It also established
a fundamental indeterminism associated with the behavior of light in interaction with matter. In Norton's words:

\begin{quote}

The master stroke of that paper comes in its sixth section. Einstein takes what looks like a
dreary fragment of the thermodynamics of heat radiation, an empirically based expression for
the entropy of a volume of high-frequency heat radiation. In a few deft inferences he converts
this expression into a simple, probabilistic formula whose unavoidable interpretation is that
the energy of radiation is spatially localized in finitely many, independent points. (Norton, 2005, 72)
\end{quote}

Let us now briefly review Einstein's argument. He assumes only a system of a finite number $N$ of independent, moving localizable points corresponding 
to some varying volume V and shows, given the additivity of entropy $S$ , that this sort of system obeys the Ideal Gas law, i.e.:

$$S - S_0 = kN \ln \frac{V}{V_0} \eqno (11) $$

Einstein thus identifies the Ideal Gas Law as a macroscopic ``signature'' (Norton's term) of microscopic statistical independence.
He then goes on to apply a similar analysis to radiation in the range conforming to Wien's Law, i.e., the high-frequency portion of the blackbody spectrum.
Modeling radiation as localized to the above system of N moving points and also subject to Wien's Law, he finds that the change in entropy of the radiation
for a fluctuation to a smaller volume is:

$$S - S_0 = k \frac{E}{h\nu}  \ln \frac{V}{V_0} \eqno (12) $$

Thus, Norton concludes, ``Therefore...a definite frequency cut of high-frequency heat
radiation carries the characteristic macroscopic signature of a system of many spatially independent components.''
These are discrete, independent systems whose probabilistic (random) aspects are ontological, not epistemic. Specifically, the probabilities
$W$ obey the relation

$$W = \frac{V}{V_0} e^N \eqno(13) $$ where $N = \frac{E}{h\nu}$. 

Norton goes on to point out that the latter applies
in general to localizable systems assumed subject to statistical independence. He emphasizes that
this law does not depend on detailed microscopic laws of motion; on the contrary it, is not derived
from deterministic classical laws:

\begin{quote}

For a century and a half, it has been traditional to introduce the ideal gas law by tracing
out in some detail the pressure resulting from collisions of individual molecules of a gas
with the walls of a containing vessel. This sort of derivation fosters the misapprehension
that the ideal gas law requires the detailed ontology of an ideal gas: tiny molecules, largely
moving uniformly in straight lines and only rarely interacting with other molecules. Thus,
it is puzzling when one first hears that the osmotic pressure of a dilute solution obeys this
same law. The molecules of solutes, even in dilute solution, are not moving uniformly in
straight lines but entering into complicated interactions with pervasive solvent molecules.
So, we wonder, why should their osmotic pressure conform to the law that governs ideal
gases?
The reason that both dilute solutions and ideals gases conform to the same law is that
their microstructures agree in the one aspect only that is needed to assure the ideal gas law:
they are both thermal systems consisting of finitely many, spatially localized, independent
components. 

\end{quote}

Thus, the Ideal Gas Law is the ``signature'' of statistical independence of N degrees of freedom confinable
to a varying volume, and it applies to any system capable of that description, including photons.
We should however add one dissent: specifically, from the aspect of the argument taken as showing that photons are localized at spacetime points. That conclusion
 comes from an assumption that a volume V can be divided without limit and still contain a well-defined amount of energy (as assumed in Dorling 1971). The idea that a well-defined amount of energy can be contained in an infinitesimally small volume contradicts the Heisenberg Uncertainty Principle $\Delta p \Delta x \ge h$, so this is revealed as a classical idealization that breaks down at the quantum level. Moreover, in any case, despite its dependence on V, the Ideal Gas partition function $Z(T)$ does not warrant a conclusion of localizability of the degrees of freedom (whether classical or quantum), since it is derivable from a quantum system in terms of its thermal de Broglie wavelength (more precisely, from considerations of the 
 energy eigenstates allowable in a box of volume V). Specifically, $Z(T) = \frac{V}{\lambda_{th}^3}$ where $\lambda_{th} = \frac{h}{2 \pi mkT}$.\footnote{Norton comments regarding (13): ``In other words, this formula tells
us that in measure one of infinitely many cases in which we might check the state of the
radiation energy E, it will be distributed in $n$ spatially localized points of energy of size $h \nu$.''. The subtlety here concerns how one would ``check'' the system's state, which amounts to a measurement. Once we do that, we localize they system, so that the localization cannot be assumed to have applied prior to the measurement.}
 
 In any case, the key finding is that under specific conditions consistent with Wien's Law, electromagnetic radiation behaves
 like a set of independent degrees of freedom not subject to deterministic laws. This result applies to radiation in a thermal context,
 and thus serves as a basis for the reality of a fundamental microscopic indeterminism in physical processes. 

\section{Szilard and Brillouin: Measurement blocks the Demon}

Before addressing the above-mentioned possibility of a genuinely Markovian component to the micro-dynamics that would serve to
underlie the notion of ``heat'' and entropy increase as derived in the H-theorem, let us return to approaches to Maxwell's Demon that explored the measurement process as as source of entropy that would block the Demon. 

\subsection{Szilard's engine} 

Leo Szilard (1929) explored this issue by modeling the Demon as using his knowledge of an atom's location in a chamber to generate work without
an entropy cost. Specifically, he envisioned four steps: (I) an atom in a chamber; (II) the placing of a partition in the chamber to divide it
half; (III) ascertaining which side the atom is on and attaching a load to the partition which is then pushed by the atom's pressure for an amount of work $W = p \Delta V$; (IV) removing
the pushed-out partition to prepare for the next cycle.  The basic set up is typically illustrated as in Figure 1.

\begin{figure}[h]	
   \centering
    \includegraphics[width=0.8\textwidth]{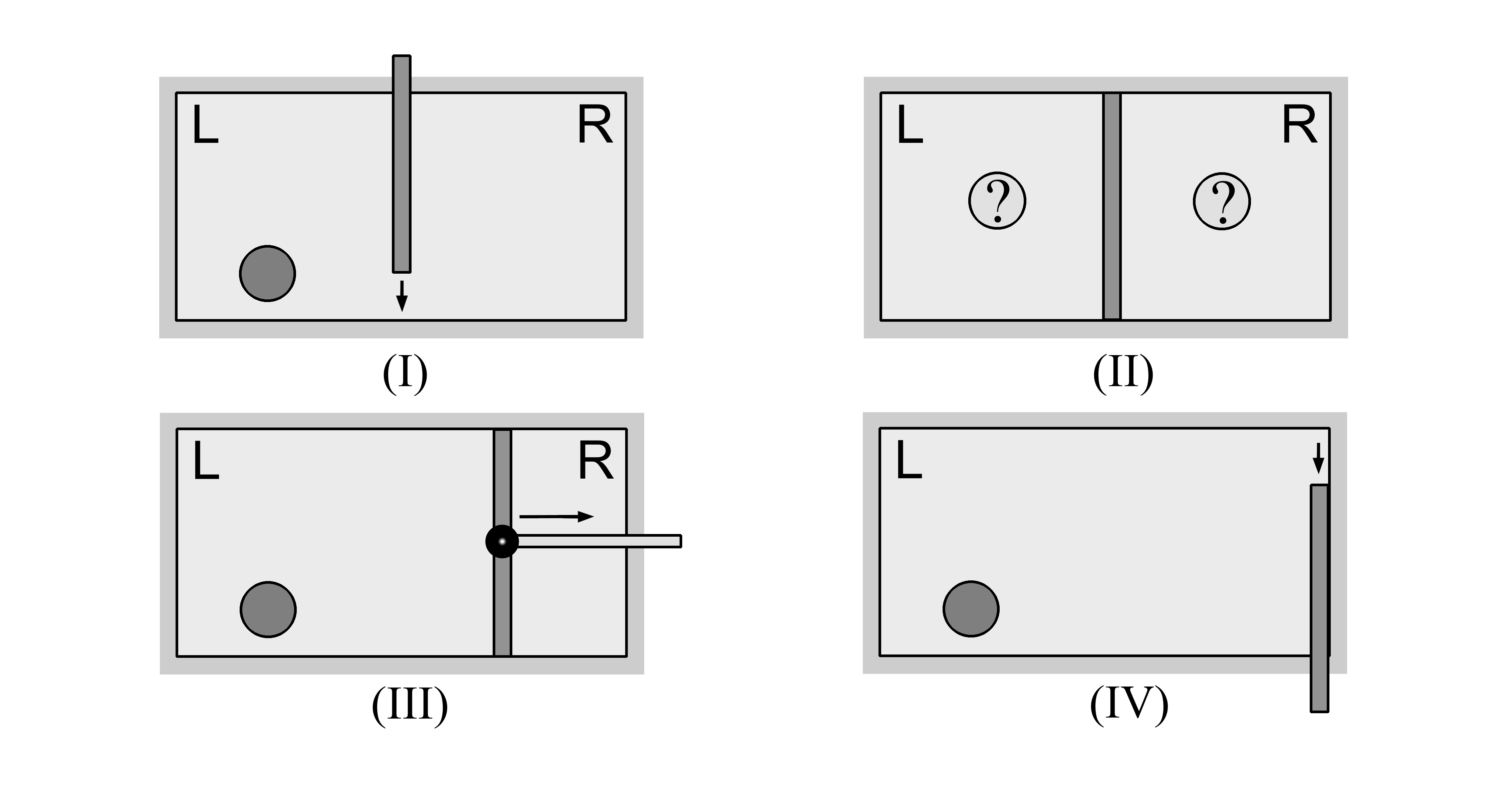}
    \caption{The basic Szilard Engine setup}
    \label{fig:your-label}
\end{figure}

Szilard did not propose a specific physical explanation for why the Demon should not be able to accomplish this apparent violation of the Second law; instead, he assumed, by reference to
the Second Law, that the Demon must incur an entropy cost through the need to interact with the system to measure the atom's position. However, at the classical level
and assuming frictionless and reversible operations, there is no clear source for the entropy increase. (We will see later, in Section 7, that it is the quantum level that
comes to the rescue of the Second Law.)

\subsection{Brillouin's argument}

L. Brillouin (1951) argued that the Demon cannot violate the Second Law based on his need to inject energy into the gas in order to
detect (measure) the molecules as required for sorting. However, his analysis was not based on a fundamental quantum
principle but rather on the need to use an amount of energy $h\nu_1 >> kT$ in order to obtain sufficient illumination
to distinguish the molecule from the background at temperature T. According to this analysis, Brillouin defines the initial
entropy of the system as $S=k \ln P_0$  where $P_0$ is the number of microstates corresponding to the macrostate of the gas
at temperature T. He then notes that injecting the above quantity of energy increases
the entropy of the gas by an amount  $\Delta S_D = \frac{h\nu_1}{T}$. He defines this quantity via the constant $kb$, noting
that $b>>1$. He characterizes this as a way for the Demon to obtain a quantity of  ``information'' $p$, which he uses to effect sorting and which decreases the microstate count to $P_0-p$ ``since the state of the system is more completely specified.''.  The 
final entropy is $S_f = k \ln (P_0 - p)$. Since $p<<P_0$, to a good approximation the change in entropy is
 $\Delta S \approx  k \Delta (\ln P)    = -k(p/P_0)$, 
so that this is generally a small decrease in $S$. Meanwhile, however,  the entropy injected by the Demon is $kb >> k (p/P_0)$
so that the Second Law remains intact. 

Brillouin's analysis can be criticized for the use of approximations that may not hold; e.g., much depends on the assumptions
concerning how large the Demon's energy expenditure must be in comparison to the reduction in entropy of the gas. In addition,
his introduction of information-related concepts such as ``negentropy'' (which he defined as a ``source of disequilibrium'') arguably needlessly complicated
the analysis without actually providing any new insights, since one can evaluate increases and decreases in entropy without such concepts. 
Brillouin's venture into information theory
was followed by numerous discussions attempting to relate the challenge of the Demon to information-theoretic concepts and
principles, among them logical notions of reversibility and irreversibility in the context of computational processes.  
However, Norton (2025)  has argued that the foray into information theory was ultimately counterproductive in the study of Maxwell's Demon.

\section{Erasure Approaches and Critiques}

As noted above, neither Szilard nor Brillouin made a clearly sound and universal case for measurement as the origin
of the entropy cost to the Demon. Against this backdrop, an alternative view emerged, pioneered by C. Bennett, that 
erasure of information was the key.

\subsection{Bennett's erasure approach}

  Bennett (1982, 2003) argued that the Demon could make the necessary measurements
on the gas molecules without entropy cost,  but would need to store the results of his measurement in a ``memory degree of freedom'' that
does not have infinite capacity and would therefore need to be periodically erased in order to make room for new information. This 
erasure was assumed to be inevitably accompanied by an entropy cost as originally suggested by Landauer (1961). Specifically,
Rolf Landauer conjectured in his (1961) that a minimal entropy cost of $k \ln 2$ is incurred for any 1-bit process implementing 
``logical irreversibility'', such as the merging of two computational paths. The latter is assumed to correspond to ``erasure,''
so that ``Landauer's Principle'' came to be known as the idea that ``erasure of information'' is always accompanied by such an
entropy cost. In this context ``information'' was taken as that general quantity defined by Shannon, i.e., $I_S = \sum_i p_i \ln p_i$
where $\{ p_i\}$ denote the probabilities of micro-configurations $\{ i \}$ (Shannon, 1948). 
 Thus, the Demon came to be analyzed in computational terms, in which context ``Landauer's Principle'' was invoked, in the form 
 of the claim that erasing a computational memory register was inevitably dissipative, thus putatively saving the Second Law.
 
 Bennett's formulation can most easily be seen in terms of the Szilard engine, the basic form of which can also be taken as modeling the Demon's
 memory degrees of freedom, which has two corresponding states R and L. Let us label the the engine atom as A and the Demon's memory storage degree of freedom as D.  D has some initial state (either R or L); depending on the result of the Demon's measurement on A, it will either stay there
 or move to represent the appropriate result. It is then assumed that the Demon discovers which side of the engine A occupies
 by simple inspection, and copies that into his memory degree of freedom D, such that ``A = R (or L)'' is mirrored in ``D=R (or L)''. 
 At this point, Bennett's Demon has seemingly evaded the entropy cost pointed to by Brillouin (1951), as discussed above, and can
 now allegedly use the information to extract work from the engine through reversible manipulations. 
 
 However, according to Bennett, the Demon must now erase and reset his memory to the initial state. This is where
 Landauer's Principle is invoked: the claim is that an irreversible expansion step is required for full erasure of the 
 memory, in the form of a removal of the partition separating the regions R and L, followed by a compression to
 reset the memory to the original state. Since this halves the phase space corresponding to D, the cost is
 an amount of entropy $S \ge k \ln 2$, which just balances the amount of work gained from the engine, thus allegedly
 saving the Second Law.
 
 While an early consensus emerged around Bennett's account  (e.g. Ladyman {\it et al}, 2007), it has also been criticized, primarily by Earman and Norton. We now turn to that dissent and similar ones from other authors.

\subsection{Dissents from the Bennettian approach}

 Earman and  Norton (1998, 1999). and Norton (e.g., 2005, 2013, 2017, 2025) have contributed numerous publications
 dissenting from the consensus that emerged around Bennett's account (e.g., 1982, 1987, 2003). Their criticism is directed at two main
 features of Bennett's formulation:
 
 1. The assumption that measurement is trivial and amounts to mere copying of pre-existing values. 
 
 2. The use of Landauer's Principle as the primary tool of ``exorcism'' of the Demon. 
 
	Firstly, concerning (1), the above authors have critiqued Bennett's claims that a Demon would be able to ascertain needed information 
without incurring an entropy cost. An example is Bennett's ``keel and key'' device (1987, 114) that would ostensibly detect the location of a gas molecule
without any dissipation. Earman and Norton (1999, 13-14) point out that thermal fluctuations would cause the device to rock uncontrollably and
thus disrupt the envisioned measurement process: 

\begin{quote}
 Two pistons are lowered in each half of the
cylinder in such a way that the pressure from the molecule will tip a delicately
balanced keel, attached to the pistons. That tipping purportedly reveals the
position of the molecule without entropy cost...The difficulty with
Bennett’s proposal is that the mechanical keel system described is an ordinary
mechanical device that would be governed by a Hamiltonian mechanics. As
a result we must presume that it would behave like a canonical thermal system.
That would mean that it would be subject to the usual fluctuation phenomena.
Intuitively, these fluctuations would arise as a wild rocking of the keel resulting
from its recoils upon each of the many impacts with the molecule of the gas. If
the keel is light enough to be raised by the pressure of the one-molecule gas, then
it must have very little inertia and such rocking is to be expected. Presumably
this wild rocking would obliterate the keel’s measuring function.
\vspace{-3 mm}

\end{quote}

Regarding (2), the above authors have noted the absence of an independent proof of Laundauer's Principle 
and argued that claims to save the Second Law from the Demon by way of Landauer's Principle amount either
to circularly invoking the Second Law or presenting individual examples in which Landauer's Principle appears to hold (as opposed
to a general proof of the latter).

Other authors have joined the basic dissent of Earman and Norton. Hemmo and Shenker (2016, Section 5) argue that erasure is not necessarily dissipative and on this basis argue that a Maxwell's Demon is not ruled out. Their analysis is based on an epistemic reading of statistical uncertainty. Maroney (2005, 2009) has critically analyzed the claimed 
connection between information and entropy assumed in the computational form of Landauer's Principle. 
R. E. Kastner and A. Schlatter have contributed publications indicating agreement with the basic objections of Earman and Norton,
and have also introduced an argument that the Demon is thwarted by quantum-level considerations including the uncertainty principle.
Their proposal is discussed in Section 7.

\section{Causation to the (partial) rescue?}

C. Weaver (2021) hands victory to the Demon, but argues that the Second law holds at a macroscopic level sufficient to ``save the appearances''
of typical observers. His main aim is to vindicate the SZA and thus defend the H-theorem from reversibility objections such as that of Loscmmidt,
despite his assumption that the laws are reversible at the micro-level. His basic proposal is to point to a fundamental notion of causation not modeled (and, he claims, not modellable) in those laws.  He terms this concept ``causationF,'' which he sees as vindicating the irreversibility of the SZA. In short, he argues
that ``causationF'' results in correlations that are time-asymmetric:

\begin{quote} ``That causation (call it fundamental causation or
causationF) drives the engine of entropic or minus-H increase. It results in correlations (that’s why
you can use correlations to find causal interactions), correlations that are one-sided precisely
because causationF is asymmetric. That is to say, obtaining causalF relations in entropy producing
collisions explain the [SZA].'' (p. 46)
\end{quote}

While the attempt to find a physical grounding for the SZA is admirable, arguably the notion of
``causationF to the rescue'' (CFR) is not only (i)  insufficiently robust to underlie the time-asymmetric entropy increase of the H-theorem,
but (ii) actually contradicts the essence of the SZA.  Concerning (i): the Demon can use that same causationF (under an assumption of dissipation-free measurement) to reduce entropy and  thereby contradict the H-theorem. While Weaver concedes that under his view the Demon can defeat the second law, he does not
appear to appreciate the inconsistency in the idea that it is causationF that supports the 2nd law in the form of the H-theorem while also
 allowing its violation for a small enough creature. In any case, regarding (ii),
attributing support for the SZA to asymmetry of the creation of correlations is a misreading of what is doing the ``work'' in the theorem,
as discussed in Section 2.2. Recall Burbury's pinpointing of the SZA assumption as one of randomness in the prediction of future 
correlations: ``For any given direction [of mutual velocity] before collision, all
directions after collision are equally probable.'' (Burbury 1894, 78). Under CausationF, a past state will be correlated with a future state,
which is explicitly denied by the SZA randomness assumption. 

Thus, there is s subtle distinction to be made in connection with the SZA. All that is actually needed to derive the Second Law is the
fundamental stochastic treatment; no explicit time asymmetry really comes into play. Specifically,  it is often stated that the irreversibility coming from the SZA
is due to the assumption that molecules are correlated after collisions rather than before, and this way of putting the situation is what Weaver has
addressed in his proposed solution. But in fact, any correlations
are wiped out from the computations at each step of the evolution of the system, since the probabilistic law applies to any transition from
microstate A to microstate B. That is, as Burbury pointed out, the irreversibility comes from the Markovian nature of the evolution, not from any assumption of
time asymmetry either explicit or implicit.\footnote{This point is reinforced by Einstein's derivation of the Ideal Gas Law through an assumption of independence of the degrees of freedom comprising the system, in particular independence of their spatial positions from their energies. This clearly requires neglect of any correlations of the molecules as a result of collisions. For a review, cf. Norton, 2006.} In fact, the transition rates themselves are time-reversible in the sense of being symmetrical between initial and final states, as we see from the Master Equation derivation
of the Second Law. 

 Thus, again it is the fundamental statistical treatment, which in the context of classical mechanics can only be ignorance of what is taken to be a microscopic fact (`the real microstate of the system at any given time')
 that is doing the heavy lifting in proving the H-theorem, while that same ignorance-based approach is what allows it to be violated (by the Demon). 
 Appealing to causation as a justification for time-asymmetry misreads what is actually doing to work in obtaining the Second Law. It is stochasticity,
 not time asymmetry of correlations, since temporal correlations (in either direction) contradict the fundamental stochastic assumption used to derive the H-theorem.

\section{Ontological Uncertainty As Key to Thwarting the Demon}

As alluded to above, the very possibility of a Demon in the first place arises from the classical treatment of entropy as based on the assumed ignorance
of a typical (macroscopic) observer who does not have epistemic access to the assumed actually determinate positions and momenta
of micro-particles undergoing supposedly deterministic (Liouville) evolution, or at least possessing determinate values of relevant
observable that are assumed to be in-principle available but not normally accessible. The Demon is the atypical observer who {\it does}
have access ``by simple inspection'' (Maxwell 1867) and who can supposedly intervene without any entropy cost (through reversible operations) so as to effect an evolution
away from equilibrium. We should note again that the Liouville evolution in phase space is a classical idealization, since quantum systems
are not describable by determinate phase space points, due to the uncertainty principle. Analyses that do take note of this point generally assume a wave packet description, in which the wave packet itself follows a deterministic evolution according to the Schrodinger equation.\footnote{However, a quantum system in a box is described by energy eigenstates that are not at all localized in phase space (i.e., not a wave packet) and whose transitions among those states are not described by continuous evolution but by the experimentally corroborated Fermi's Golden Rule, as noted above; we address this counterexample in what follows.}  In that case, however, there remains no ``randomness,'' i.e., no fundamental indeterminism of the evolution, that is the hallmark of thermal processes. Thus, most extant analyses continue to treat ``heat'' as something arising from the ignorance of an observer who is not otherwise privy to the assumed actual deterministic trajectories of what is conceived of as either a phase space point or a phase space minimal area ~ $h$.  Against this backdrop, the Demon remains a possibility, and it is to their credit that Hemmo and Shenker acknowledge this point. 

However, in this section, we will lay out an argument that disagrees with the affirmative response of Hemmo and Shenker  to the question "Is Maxwell's Demon compatible with the physical laws of our world?"  It takes up the following challenge by these authors:

\begin{quote}
According to the conventional wisdom, Maxwellian Demons do not exist in our world as a matter of fact. There are several ways of defending this position against our argument. The first and obvious reply would be that our construction of the Demon is wrong in that it contradicts some principle of classical or quantum mechanics. This alleged principle should rule out the sort of harmony between the dynamics and the partition to macrostates required by our construction of the Demon. However, we believe that our construction does not contradict any principle of mechanics (classical or quantum).
\end{quote}

The present analysis points out that in fact there is a principle that is violated by their analysis: namely, the Heisenberg Uncertainty Principle (HUP).  The HUP precludes the localization of a degree of freedom without an energy cost (due to the increasing spread of momentum of the system). On the other hand, it also precludes an exact determination of 4-momentum without incurring a corresponding uncertainty in spatiotemporal parameters.  It will turn out that either of these features serves to foil a Maxwell's Demon (depending on how the Demon is modeled). Before proceeding with the details, however, we should distinguish two notions of uncertainty that are both required in analyzing the statistical treatment of thermodynamics:\smallskip

1. Physically random processes (dynamic uncertainty concerning state transitions--``damned quantum jumps'')

2. Uncertainty in system properties (Heisenberg uncertainty principle, $\Delta x \Delta p \ge 0; \Delta t \Delta E \ge 0$).
\smallskip

While we need uncertainty type (1) for a valid derivation of the Second Law as argued above, it will turn out that we also need type (2)
to foil a Demon's attempt to violate the law. In what follows, we review a proposal of Kastner (2025) that lays out 
the barrier to the Demon that arises from the quantum uncertainty principle. 

\subsection{Heisenberg Uncertainty Principle vs. Demon}

In this analysis we note that real physical systems, such as gas molecules, are quantum systems. In this context, neither an ordinary experimenter
nor a Demon are merely ``ascertaining'' properties in the sense of remedying their ignorance about pre-existing values of observables, as is assumed
in Maxwell's thought experiment. The latter assumption, that measurement is mere copying of some pre-existing value, is assumed by Bennett
and his followers. Yet it is explicitly contradicted by the uncertainty principle . Instead, at the quantum level, we are creating such values (or at least bringing them about) via measurement. If the uncertainties $\Delta x \Delta p \ge 0; \Delta t \Delta E \ge 0$  reflect ontological indeterminacy, which would appear to be the case
for quantum systems, then the ``information'' does not characterize the ignorance of an observer (as is inevitable in classical statistical mechanics) but rather genuine ontic uncertainty concerning the system's properties.\footnote{The reader might object that quantum theory can also be cast in a form utilizing hidden variables, which are viewed as determinate properties. However, these are not classically determinate; for example, the DeBroglie-Bohm pilot wave theory takes position uncertainty as epistemic, but the strong coupling of those positions to the quantum state (via the "quantum potential") makes the uncertainty based on the state ontologically consequential. Thus, putative Bohmian corpuscles remain subject to the uncertainty principle.} 

Before proceeding to that formulation, some definitional distinctions are crucial. Since we are now dealing with the concept of information, we need to make precise what is meant by that term in the following analysis. As Shannon (1948) showed,
one can indeed quantify the {\it informational resources} one is dealing with in any particular process, such a a signal communicated through a set of bits
(units with two degrees of freedom).  This notion of ``information'' is a very general concept that quantifies such systems or processes, and it turns out to have the same form
as Boltzmann's expression for entropy; namely, $I = -\sum_{i} p_i \ln p_i$ where $p_i$  is the probability of microstate $i$. So for example, in a system with one bit,
$i = \{1,2\}$, for $p_i = 1/2$ 
the information content is $ I(bit) = \ln 2$. These ``microstates'', however, are a generalization from the microstates appearing in thermodynamic quantities,
and one cannot in general identify {\it thermodynamic} entropy $S$ with (Boltzmann's constant times) the quantity $I$. Despite this fact, the tradition of calling Shannon's quantity ``entropy''
has persisted, and arguably contributed to confusion surrounding the nature of the Second Law and its challenges, such as that of Maxwell's Demon.  

In the analysis that follows (presented in Schlatter and Kastner, 2024; Kastner and Schlatter, 2024) and Kastner, 2025), various forms of information are physically relevant, but only one is identified with thermodynamic entropy: the form that involves energy, as in keeping with the basic definition $\Delta S = \Delta Q/T$ and the observations in Section 3. What we find is that reduction of uncertainty in pointer-type information (such as the whereabouts of the molecule in the Szilard engine) is accompanied by an entropy cost, as quantified by the informationally quantified form of the Heisenberg Uncertainty Principle (HUP). Conversely, one also finds that reduction in entropy is accompanied by an increase in location uncertainty, which prevents sorting such as that required for the Demon's operation in Maxwell's original scenario. 
We now present the relevant form of the HUP, which for specificity we can call the  {\it Quantum Information Uncertainty Relation} (QIUR). It is obtained by casting the HUP into a form reflecting the information content associated with wave function uncertainties:

$$I(x,p) = - \int_{-\infty}^{\infty} |\psi(x)|^2 \ln  |\psi(x)|^2 dx -  \int_{-\infty}^{\infty} |\phi(p)|^2 \ln  |\phi(p)|^2 dp \ge \ln (he/2)\eqno(14)  $$

\noindent where $\psi(x)$ is the position wavefunction and $\phi(p)$ is the conjugate momentum wavefunction.\footnote{The factor of $h$ in the right-hand side is needed because
the standard deviation of the wavefunctions are expressed in units of $h$.}  The above expression for the joint information for Fourier-transform pairs was presented by Hirschman (1957) and Leipnik (1959), both of whom were elaborating the pioneering work of Weyl (1928).  Now, it becomes evident that the uncertainty principle has real physical consequences for the thermodynamic entropy $S$ of systems when we identify the latter as the information associated with the momentum state, i.e.:

$$ S = k I(p) = -k  \int_{-\infty}^{\infty} |\phi(p)|^2 \ln  |\phi(p)|^2 dp \eqno(15)$$

For Gaussian wavefunctions, we have equality, and any reduction in uncertainty--i,.e, reduction in ``information'' $I$--of either parameter, $x$ or $p$, results
in a corresponding increase in the information $I$ of the conjugate parameter. Thus, if a measurement localizes the atom (reducing its position-information) in the Szilard engine, its entropy $S = k I(p)$ increases by the same amount. For wavefunctions other than a Gaussian, the resulting increase is larger. Conversely, a reduction in $S$ through 
a measurement of momentum results in a corresponding increase in the position-information $I(x)$, and this is what prevents the original Maxwell Demon
from sorting gas molecules according to the speeds and placing them on the appropriate side of the container. Specifically, determining their speeds delocalizes them so that they cannot be ``put through a door'' as envisioned in the classical scenario.

We now calculate the specific costs associated with each type of measurement to show the above quantitatively. For this purpose, the easiest approach
is to use the expression for $I(x)$ in terms of the variance of the probability distribution $\sigma_{x}^2$ corresponding to a given wavefunction $\psi(x)$ :

$$ I(\sigma_x) = \frac{1}{2} \ln (2 \pi \sigma_{x}^2 e) \eqno(16) $$

The above is a general relation that holds for any parameter $x$ (cf. Leipnik 1959). 

\subsection{The Szilard Engine Demon}

Now let us consider, for simplicity, the ``molecule in a box'' form of Szilard's engine as schematically depicted in Figure 1. (We use a one-dimensional box for simplicity without
loss of generality.)  While the exact solutions for
this situation are energy eigenstates, for simplicity let us assume that due to decoherence the wavefunction can be modeled as a Gaussian
to a good approximation. This will instantiate the minimum bound (equality) for equation (14). (Any other wavefunction subject to a reduction in $I$ will yield a larger increase
in the $I$ of its conjugate wavefunction.) For a box of length L, the molecule's initial position variance is of the order of the box, i.e.: $\sigma_{x,i}^2 \approx L^2$.  For a Gaussian, the initial variance associated with conjugate momentum wavefunction is  $\sigma_{p,i}^2 \approx \frac{h^2}{4L^2}$.  After the Demon places the partition in the center, the position uncertainty is reduced by 1/2, so the final position variance is $\sigma_{x,f}^2 \approx \frac{L^2}{4}$. Meanwhile, the momentum variance increases to
 $\sigma_{p,f}^2 \approx \frac{h^2}{L^2}$. 
  The difference in initial and final entropies of the molecule can be found using the definition (15) and the relation of the variance to information (16), i.e.:

$$\Delta S = S_f - S_i = \frac{k}{2} \ln \frac{\sigma_{p,f}^2}{\sigma_{p,i}^2} = \frac{k}{2} \ln 4 =  k \ln 2 \eqno(17) $$

 We immediately see that (17) matches the entropy cost traditionally associated with Landauer's Principle, and this is no coincidence, 
 as we will discuss further in Section 7.4. Equation (17) tells us that the entropy cost of inserting the partition, and thus decreasing the position uncertainty of the molecule's wavefunction,
results in an entropy increase that exactly compensates the decrease in entropy resulting from reducing the volume accessible to the molecule. 
Of course, any work obtained from that reduction in volume by the Demon is also thus paid for by the same entropy increase.  The key point
is that this entropy cost  is incurred {\it regardless of whether anyone (including a Demon) learns which side the atom is on after insertion
of the partition}. Thus, entropy is not about knowledge, despite the long (classical) tradition of assuming this to be so.  Entropy is a measure
of the energy transferred into or out of a given system (at temperature T) in any particular processes, as given in Clausius' original definition. And as noted in Section 2.3,
the terms ``heat'' and ``work'' for energy are interchangeable in the sense that both involve path-dependent energy transfer.
 The Demon had to place the partition such that it would result in a decrease in the molecule's position uncertainty.
 That resulted in a corresponding increase in the molecule's momentum uncertainty. The latter reflects an increase in the heat content of the system,
 since there must be enough energy available to support the larger momenta reflected in the increase of $\sigma_p$. That energy came from
 the Demon's act of inserting the partition.

 Now, under the prevailing classical and pseudo-classical conceptualizations, it is hard to see why the mere placing of
 a partition could constitute work on the molecule. Indeed, the usual depiction of this situation suggests that one is
 simply applying a constraint that does no work on the system, since it is being placed ``where the atom is not'', as in Step I 
 from Figure 1:
 
\begin{figure}[h]	
   \centering
    \includegraphics[width=0.3\textwidth]{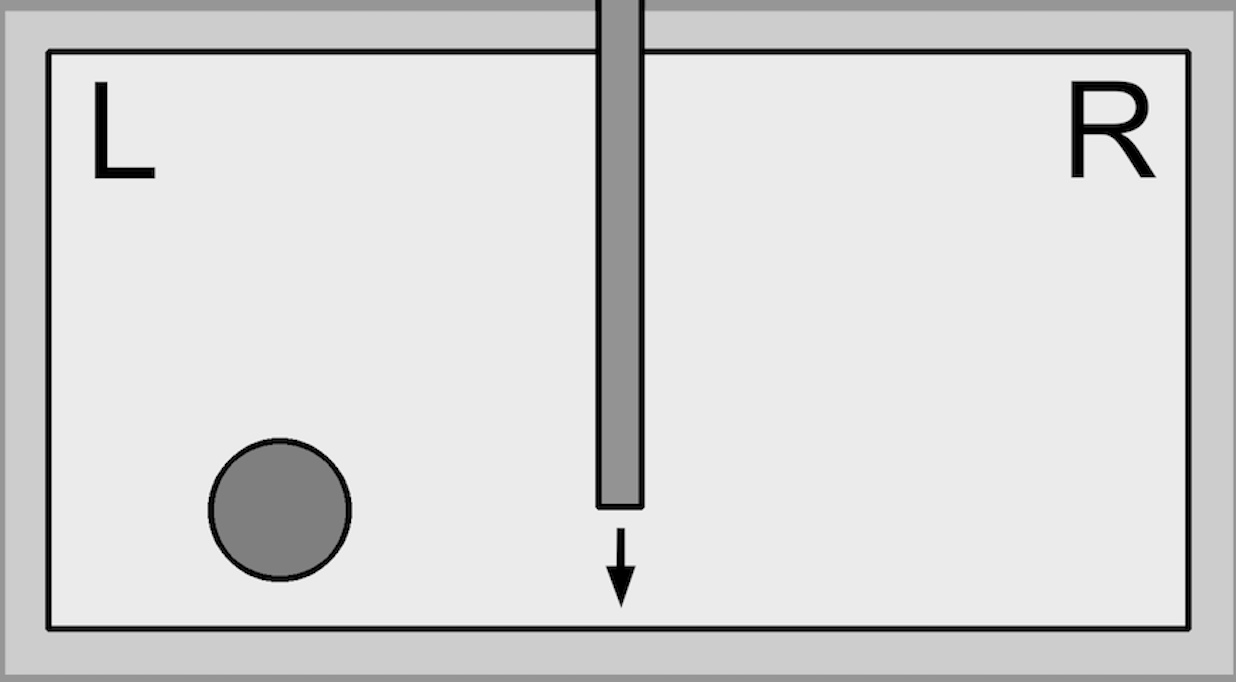}
    \caption{Gas molecule depicted as a localized classical particle}
    \label{fig:your-label}
\end{figure}

But this is where the classical idealization breaks down, since the atom is indeed a quantum system and its most accurate
description is an energy eigenstate. Even in the ``classical limit'', at high temperatures, the atom would be in a mixed state
corresponding to the Maxwell-Boltzmann distribution, which means that it is still delocalized, as in Figure 3.

\begin{figure}[h]	
   \centering
    \includegraphics[width=0.5\textwidth]{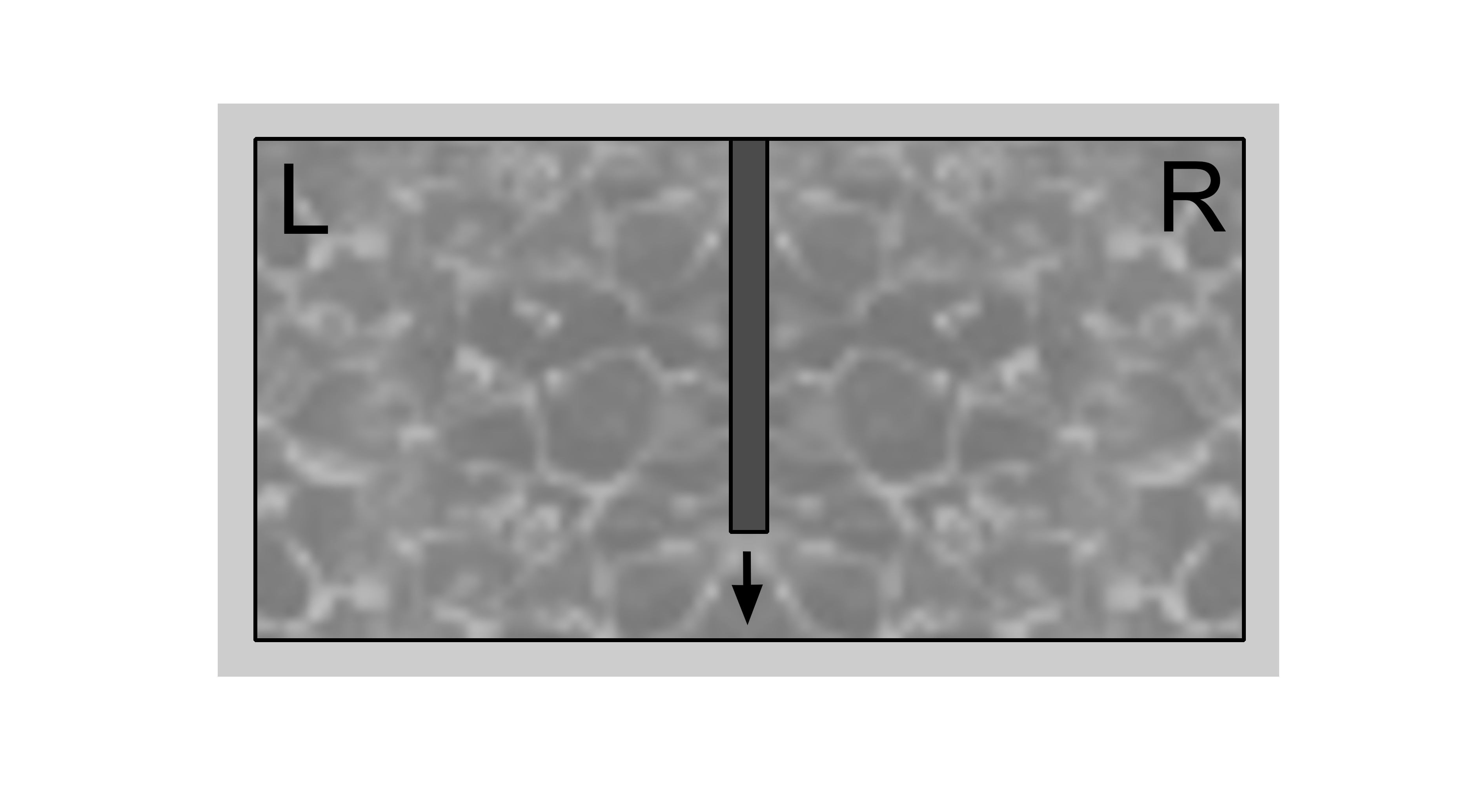}
    \caption{A real, delocalized quantum molecule}
    \label{fig:your-label}
\end{figure}

Thus the placing of the partition is indeed crucially affecting its quantum state by confining it to a smaller volume. 
In effect, the partition insertion functions as a quantum measurement, where here the term ``measurement'' means
an interaction resulting in a localization (reduction in uncertainty) of the system's wavefunction $\psi(x)$. Specifically, the molecule's
 wavefunction is altered to either a narrow left-hand state $\psi_L(x)$ or a narrow right-hand state
$\psi_R(x)$. It does not matter whether or not any external party (Demon or other) knows which one it is; the 
entropy cost has already been incurred.

 To appreciate this point, we must disambiguate the term ``measurement''  from epistemological considerations; i.e., considerations of the state of knowledge of an entity outside the system itself (whether a Demon or ordinary experimenter). It does not matter on which side of the container the molecule ends up after the partition insertion (measurement); its entropy has already increased due to relation (14). Thus, the Demon is foiled by the quantum measurement process, but not through epistemological considerations associated with his having to find out where the molecule is, as is conventionally assumed. Instead, it is the intervention of partition insertion that forces localization of the molecule, and the associated uncertainty principle, that yields the entropy increase. In this light, we see that the efforts by Bennett and his followers to argue that measurement can be carried
out without an entropy cost, such as via his ``keel and key'' device (Bennett, 1988)\footnote{Although Norton, 2017 argues that it would not work as claimed.}
 are solely focused on the epistemological aspect--i.e., the Demon's state of knowledge--rather than on the actual
act of quantum measurement, which for the Szilard engine is what affects the particle's state in such a way as to increase its entropy.\footnote{This oversight is repeated even in treatments seeking to address the quantum level; e.g., Yavilberg (2025), in which it is stated that ``when the [partition] is inserted the measurement is not performed yet.'' This assessment clearly overlooks the relevant quantum states and Heisenberg uncertainty principle.}

\subsection{The Maxwell ``Speed-Demon''}

We now turn to the converse case, i.e., the original Maxwell Demon, who engaged in sorting based on measuring molecular speeds
rather than the position of a molecule. The ``Speed-Demon'' is subject to (14) in a different way, but one that still foils his ability to sort, as follows.
The Demon is assumed to be measuring speeds, so that he is reducing the uncertainty in $\phi(p)$. Thus the uncertainty in each molecule's position  $\psi(x)$
must increase accordingly. However, in order to get a molecule through the opening between the chambers, it must fit through the door, which
has to be sufficiently small to prevent leakage. To quantify, this size $d$ can be no larger than $d \approx \frac{h}{4 \pi \sigma_p(p_{rms})}$, 
where root-mean-square momentum $p_{rms} \approx \sqrt{3mkT}$. However, in order to measure momentum accurately, one needs a low-energy
photon, i.e. $h \nu_{low} \ll kT$. This results in a correspondingly reduced post-measurement $\sigma_p \approx \sqrt{3mh\nu_{low}}$, i.e., a decrease
in the ``local entropy'' of that single molecule. However, due to the conjugate
relationship between $x$ and $p$, for a Gaussian wavefunction that implies  $\sigma_x \approx  \frac{h}{4 \pi \sqrt{3m h \nu_{low}}} \gg d$; i.e. the molecule's
position-information $I(x)$ greatly increases accordingly. Any
other wavefunction will result in a larger increase in $\sigma_x$ and thus $I(x)$.  This has real physical consequences in that the required momentum measurement delocalizes the molecule such that it cannot fit through the door, even if if its general location were known. The conclusion is that the ``Speed-Demon'' envisioned by Maxwell cannot
even begin to sort in the first place, once it is taken into account that real gas molecules are quantum systems. The Demon cannot reduce the entropy of the whole
gas system. All he can do is reduce the entropy of a single molecule by measuring it, but then he cannot use that measurement to effect sorting. All he has done
is inject a small amount of energy $h \nu_{low}$ into the gas that has gone, not into any efficacious sorting, but instead into delocalizing one of its N molecules.

\subsection{Landauer's Principle Gains a Physical Foundation from Quantum Measurement.}

The above results refute the fundamental assumption of the Bennettian tradition that the Demon's measurement process can be modeled as mere copying of some outcome-state that itself was arrived at without entropy cost. In short, that tradition was arrived at by neglecting the quantum uncertainty principle. Rather than having to appeal to Landauer's Principle (LP) in a presumed necessary memory erasure process, instead one can trace LP to the complementary relationship of momentum and position, where increased momentum uncertainty is the physical source of entropy cost in measurement. This is in contrast to extant arguments that either justify LP through the Second Law (making its use in exorcising the Demon circular) or through inappropriately taking particular experimental arrangements and analyses as universal. In the former case, LP is simply a restatement of the Second Law, while in the latter case it is relegated to a conjecture subject to refutation by counterexamples (as provided, e.g. in Earman and Norton 1999, 16-17).\footnote{In this discussion, the authors argue that a memory ``reset'' can be done without incurring an entropy cost amounting to the Landauer bound. The authors view this reset as distinct from ``erasure'', although it has the same effect of preparing the Demon's memory for another cycle, which is viewed in the Bennettian tradition as the key requirement.} We thus find that LP attains an independent physical basis not from conjectural arguments concerning the alleged need for erasure in computational processes, but from the uncertainty principle. In the latter context, it reappears not as a statement about computation but about the actual physics of measurement. Here, ``erasure of information'' becomes narrowing of position-related uncertainties and ``dissipation'' is the resulting increase in energy-momentum uncertainties. Thus, the above analysis (perhaps ironically) strengthens Landauer's Principle as a stand-alone result but removes its utility for computational arguments concerning  memory registers undergoing logical processes, which are classical in nature. LP is a quantum-level principle. 

What about claims in the literature to have experimentally confirmed Landauer's Principle? Firstly, of course LP does hold in the above sense, as the entropic cost of ``erasing'' position information understood in terms of definition (16).  What founders is sole identification of the LP entropy cost with general informational and logical considerations, such as ``merging of computational paths.''  In particular, what does not appear to be independently experimentally confirmed is an entropy cost of ``memory erasure,'' despite claims that this has been done.  An example of such a claim, in Lutz and Ciliberto (2015) is discussed in Kastner (2025). It is found that the experiment  they cite, by Raizen (2009), actually demonstrates an entropy cost of quantum measurement, consistent with the above analysis, despite the fact that the authors portray the entropy cost as arising from memory erasure. The authors say: “The proper resolution of the {Demon} paradox wouldn’t come for another 115 years”, referring to
the conventional account invoking memory erasure. However, in their cited experiment, the entropy cost comes from the measurement step, which is clear even from their own discussion of it:

\begin{quote}

“In the [experiment of Raizen, 2009], the optical potential serves as the demon. If an atom is determined to be moving from right to left—that is, if it encounters the excitation beam first, the trapdoor is opened. For all other atoms, the trapdoor is closed. Information about the position and internal state of the atoms is stored in the photons scattered by the atoms. Each time an atom scatters a photon, the entropy of the optical beam increases, because a photon that was propagating coherently with the beam gets scattered in an uncertain direction”.

\end{quote}

Despite the authors' portrayal of the photon beam as ``storing information,'' what they actually describe is an entropy increase in the photon beam as a result of its measuring (detecting) the atoms. Thus, the Raizen experiment confirms the analysis of this section rather than the computational form of LP: the entropy increase
comes from the measurement of the atoms by the photon beam, not from the erasure of any already-stored information. 

\section{Conclusion}

This chapter has provided an overview of the long history of debate around Maxwell's tantalizing challenge in the form of a ``light-fingered being,'' the Demon, who he proposed could violate the Second Law of Thermodynamics. We have reviewed some key developments in this long debate. While the earliest efforts to ``exorcise'' the Demon, primarily by Szilard and Brillouin, focused on measurement, those were viewed as insufficiently convincing, and arguably for good reason. Szilard basically assumed the Second Law without explaining why measurement must adhere to it, while Brillouin's account was insufficiently general and, in addition, neglected a key aspect of the quantum nature of matter, the Uncertainty Principle. Against this backdrop, and a growing trend favoring the casting of physical processes into information-theoretic terms, Bennett's appeal to erasure and Landauer's Principle gained popularity (although having its fair share of critics; e.g., Earman and Norton, Hemmo and Shenker, Maroney, Schlatter and Kastner). 

The present discussion emphasizes the importance of fundamental (ontological) indeterminism as a crucial grounding for the Second Law, noting that such fundamental indeterminism cannot be found in classical mechanics but is only available at the quantum level. This indeterminism takes two key forms: (1) discontinuous and in-principle unpredictable ``quantum jumps,'' which serve to justify Boltzmann's ``Sto\ss ahlansatz,'' and (ii) the intrinsic uncertainty in complementary system properties, as quantified by the Quantum Information Uncertainty Principle. It is shown that it is primarily (ii) that thwarts the Demon from achieving his goal in view of the unavoidable entropy cost attending quantum measurement (or, for measurements of energy-related quantities, the unavoidable disturbance of the system so as to prevent its micro-manipulation). It is further noted that ``measurement'' at the quantum level needs to be considered a physical process independent of epistemic considerations; i.e., that an entropy cost is incurred whether or not a particular agent ever learns the result of their measurement.\smallskip

In conclusion, the Second Law is upheld by quantum mechanics, which also serves to defeat Maxwell's Demon.

\newpage

References\bigskip


Bell, J.S. (1987). ``Are there quantum jumps?'' In Kilmister, Ed. {\it Schrodinger: Century of a Polymath}, p. 41. Cambridge: Cambridge University Press.


Bennett, C. (1973) “Logical Reversibility of Computation,” IBM Journal of Research and Development 17, 525–532.

Bennett, C. (1982) “The Thermodynamics of Computation: A Review,” International Journal of Theoretical Physics 21, 905–940.

Bennett, C. (1987) ``Demons, Engines, and the Second Law,'' {\it Scientific American 257}, No. 5, pp. 108-117.

Bennett, C. (1988). "Notes on the history of reversible computation," IBM. J. Res. Dev. 32, no. 1, 16-23. Jan. 1988

Bennett, C. (2003) “Notes on Landauer’s Principle, Reversible Computation, and Maxwell’s Demon,” Studies in History and Philosophy of Modern Physics 34(3), 501–510.


Boltzmann, L. (1872). Weitere Studien ¨uber das W¨armegleichgewicht unter
Gasmolek¨un, Sitzungberichte der Akademie der Wissenschaften zu Wien, mathematisch-
naturwissenschaftliche Klasse, 66, 275-370

Brillouin, L. (1951) Maxwell’s Demon Cannot Operate: Information and Entropy. I. J. Appl. Phys. 22, 334–337. 

Brillouin, L. (1962) Science and Information Theory. London: Academic Press.

Burbury, S. (1894). Boltzmann's Minimum Function. Nature 51, 78  https://doi.org/10.1038/051078a0 


Clausius, R. (1867). {\it The Mechanical Theory of Heat}s. London: Taylor and Francis.  

Dorling J. (1971), “Einstein’s Introduction of Photons: Argument by Analogy or Deduction Form the Phenomena?”British Journal for the Philosophy of Science 22,1–8.


Earman, J. (2006) “The Past Hypothesis: Not Even False,” Studies in History and Philosophy of Modern Physics 37, 399–430.

Earman J. and Norton J. (1998) “Exorcist XIV: The Wrath of Maxwell’s Demon. Part I. From Maxwell to Szilard,” Studies in History and Philosophy of Modern Physics 29(4), 435–471.

Earman J. and Norton J. (1999) “Exorcist XIV: The Wrath of Maxwell’s Demon. Part II. From Szilard to Landauer and Beyond,” Studies in History and Philosophy of Modern Physics 30(1), 1–40.

Eddington, A. (1935) The Nature of the Physical World. London: Everyman’s Library.

Ehrenfest, P. and Ehrenfest, T. (1912) “The Conceptual Foundations of the Statistical Approach in Mechanics”, Leipzig, 1912; New York: Dover, 1990.

Einstein, A. (1905a). "Über die von der molekularkinetischen Theorie der Wärme geforderte Bewegung von in ruhenden Flüssigkeiten suspendierten Teilchen". Annalen der Physik (in German). 322 (8): 549–560

Einstein, A. (1905b). Uber einen die Erzeugung and Verwandlung des Lichtes betreffenden heuristischen
Gesichtspunkt. Annalen der Physik, 17, 132–148 (Papers, Vol. 2, Doc. 14).

Einstein, A. (1916). Physikalische Gesellschaft Zurich. Mitteilungen 18 (1916): 47-62.

Einstein, A. (1970) “Autobiographical Notes,” in P. A. Schilpp (ed.), Albert Einstein: Philosopher-Scientist, vol. 2, Cambridge: Cambridge University Press.


Feynman, R. (1963) The Feynman Lectures on Physics, Redwood City, CA: Addison Wesley.

Fick A (1855). "Uber Diffusion". Annalen der Physik (in German). 94 (1): 59–86. 

Frisch, R. (1933) Experimenteller Nachweis des Einsteinschen Strahlungs-rucksto\ss es. Zeitschrift f¨ur Physik, 86, 42-48.








Hemmo, M. and Shenker, O. (2016)  https://doi.org/10.1093/oxfordhb/9780199935314.013.63 

Hirschman, I.I. (1957). "A note on entropy." Am. J. Math.  79, 152–156

Kastner, R.E. (2017) ``Quantum Collapse as a Basis for the Second Law of Thermodynamics.'' Entropy 19, 106. 

Kastner, R. E. (2025) ``Maxwell’s Demon Is Foiled by the Entropy Cost of Measurement, Not Erasure,'' {\it Foundations} 5(2), 16. https://www.mdpi.com/2673-9321/5/2/16

Kastner, R. and Schlatter, A. (2024).  "Entropy Cost of Erasure in Physically Irreversible Processes," Mathematics 2024, 12(2), 206.

Knott, Cargill Gilston (1911). Life and Scientific work of Peter Guthrie Tait, Cambridge: Cambridge University Press.


Ladyman, J., Presnell, S.,  Short, A.J.  and Groisman, B. (2007). “The connection between logical and thermodynamic irreversibility,” Studies In History and Philosophy of Science Part B: Studies In History and Philosophy of Modern Physics 38, 58–79. 

Lamb, W. and Retherford, R. (1947). ``Fine Structure of the Hydrogen Atom by a Microwave Method,'' Phys. Rev. 72, 241.

Landau, L. D. and Lifshitz, E. M. (1980) Statistical Physics Part 1, Course in Theoretical Physics vol. 5. 3rd ed. Trans: J. B. Sykes and M. J. Kearsley. Oxford: Butterworth-Heinemann.

Landauer, R. (1961) “Irreversibility and Heat Generation in the Computing Process,” IBM Journal of Research and Development 3, 183–191.

Lanford, O. E. (1976). ``On the derivation of the Boltzmann equation,'' Asterisque 40, 117-137.

Leff, H. S. and Rex, A. (2003) Maxwell’s Demon 2: Entropy, Classical and Quantum Inforamtion, Computing, Bristol, UK: Institute of Physics Publishing.

Leipnik, R. (1959). Entropy and the Uncertainty Principle. Inf. Control 2, 64–79.

Lindblad, G. (1976). Commun. Math. Phys. 48 (1976) 119

Loschmidt, J., 1876/1877, “Über die Zustand des Wärmegleichgewichtes eines Systems von Körpern mit Rücksicht auf die Schwerkraft”, Wiener Berichte, 73: 128, 366 (1876); 75: 287; 76: 209 (1877).

Lutz, E.; Ciliberto, S. Information: From Maxwell’s demon to Landauer’s eraser. Physics Today 2015, 68, 30–35. [Google Scholar] [CrossRef]
Raizen, M.G. Comprehensive control of atomic motion. Science 2009, 324, 1403. 

Maroney, O. (2005). “The (absence of a) relationship between thermodynamic and logical reversibility,” Studies in History and Philosophy of Science Part B: Studies in History and Philosophy of Modern Physics 36, 355–374 (2005). 

Maroney, O. (2009) “Information Processing and Thermodynamic Entropy,” The Stanford Encyclopedia of Philosophy (Winter 2008 Edition), Edward N. Zalta (ed.), <http://plato.stanford.edu/entries/information-entropy/>

Mellor, D. H. (2005) Probability: A Philosophical Introduction, London: Routledge.

Mielnik, B. (2017)  ``Empty Bottle: The Revenge of Schrodinger's Cat", J. Phys.: Conf. Ser. 839 012006

Norton, J.D. (2005). Eaters of the lotus: Landauer’s principle and the return of Maxwell’s demon. Stud. Hist. Philos. Sci. Part B Stud. Hist. Philos. Mod. Phys. 36, 375–411.

Norton, J. (2013). All Shook Up: Fluctuations, Maxwell’s Demon and the Thermodynamics of Computation. Entropy, 15, 4432–4483. 

Norton, J. D. (2016) "Maxwell's Demon Does Not Compute," preprint, https://philsci-archive.pitt.edu/14808/.

Norton, J.D. (2017). Thermodynamically Reversible Processes in Statistical Physics. Am. J. Phys. 2, 85, 135–145. 

Norton, J. D. (2025). "Too Good to Be True: Entropy, Information, Computation and the Uninformed Thermodynamics of Erasure." Preprint, https://sites.pitt.edu/~jdnorton/papers/Erasure\_PSA\_2024.pdf

Norton, J. D. (2006). ``Atoms, entropy, quanta: Einstein’s miraculous argument of 1905,'' {\it Stud. Hist. Philos. Mod. Phys. 37}: 71-100.

Raizen, M.G. Comprehensive control of atomic motion. Science 2009, 324, 1403. 

Schrodinger, E. (1929). {\it Science, Theory, and Man.} (German edition; English translation published 1957).

Schrodinger, E. (1957). {\it Statistical Thermodynamics}. Cambridge: Cambridge University Press.  

Schlatter, A., Kastner, R.E. (2024) A model of entropy production. Sci Rep 14, 30853. https://doi.org/10.1038/s41598-024-81671-w.

Shannon, Claude E. (July 1948). "A Mathematical Theory of Communication". Bell System Technical Journal. 27 (3): 379–423. Bibcode:1948BSTJ...27..379S. doi:10.1002/j.1538-7305.1948.tb01338.x. 

Straumann, Norbert (2017). "Einstein in 1916: On the Quantum Theory of Radiation”, preprint. https://arxiv.org/pdf/1703.08176.

Szilard, L. (1929) “On the Decrease of Entropy of a Thermodynamic System by the Intervention of an Intelligent Being,” in H. S. Leff and A. Rex (eds.), Maxwell’s Demon 2: Entropy, Classical and Quantum Information, Computing. Bristol, UK: Institute of Physics Publishing, 2003, pp. 110–119.

Uffink, J. (2001) “Bluff Your Way in the Second Law of Thermodynamics,” Studies in History and Philosophy of Modern Physics 32, 305–394.

Uffink, J. (2007) “Compendium to the Foundations of Classical Statistical Physics,” in J. Butterfield and J. Earman (eds.), Handbook for the Philosophy of Physics, Part B, pp. 923–1074.

Uffink, J. (2024) "Boltzmann’s Work in Statistical Physics", The Stanford Encyclopedia of Philosophy (Winter 2024 Edition), Edward N. Zalta \& Uri Nodelman (eds.), URL = <https://plato.stanford.edu/archives/win2024/entries/statphys-Boltzmann/>. Accessed 2/28/26.

Uffink J. and Valente, G. (2010) “Time’s Arrow and Lanford’s Theorem,” Seminaire Poincare XV Le Temps, 141–173.

Uffink J. and Valente, G. (2015) “Lanford’s Theorem and the Emergence of Irreversibility,” Foundations of Physics 45, 404–438.


Weaver, C.G. In Praise of Clausius Entropy: Reassessing the Foundations of Boltzmannian Statistical Mechanics. Found Phys 51, 59 (2021). https://doi.org/10.1007/

Weyl, H. (1928) Theory of Groups and Quantum Mechanics; Dutton: New York, NY, USA; Volume 77, pp. 393–394. 


 Yavilberg, K. (2025). "Szilard’s Engine" (Accessed 3/4/2025) https://physics.bgu.ac.il/~dcohen/courses/QuantumTopics/Contributions/Szilard\_Engine.pdf


\bigskip

Further Readings \medskip

Ainsworth, P. M. (2011) “What Chains Does Liouville’s Theorem Put on Maxwell’s Demon?” Philosophy of Science 78, 149–164.

Frigg, R. (2008) “A Field Guide to Recent Work on the Foundations of Statistical Mechanics,” in D. Rickles (ed.), The Ashgate Companion to Contemporary Philosophy of Physics. London: Ashgate, 2008, pp. 99–196.

Hemmo, M. and Shenker, O. (2010) “Maxwell’s Demon,” The Journal of Philosophy 107, 389–411.

Hemmo, M. and Shenker, O. (2012) The Road to Maxwell’s Demon. Cambridge: Cambridge University Press.

Hemmo, M. and Shenker, O. (2015) “Quantum Statistical Mechanics and Classical Ignorance,” forthcoming in Studies in History and Philosophy of Modern Physics.

Jauch, J. M. and Baron, J. G. (1972) “Entropy, Information and Szilard’s paradox,” Helvetica Physica Acta 45, 220–232.

Lebowitz, J. (1993) “Boltzmann’s Entropy and Time’s Arrow,” Physics Today, September 1993, 32–38.

von Plato, J. (1994) Creating Modern Probability, Cambridge, UK: Cambridge University Press.

Scully, M. (2001) “Extracting Work from a Single Thermal Bath via Quantum Negentropy,” Physical Review Letters 87(22), 220601.

Scully, M. O., Zubairy, M. S., Agarwal, G. S., and Walther, H. (2003) “Extracting Work from a Single Heat Bath via Vanishing Quantum Coherence,” Science 299, 862–864.

Sklar, L. (1993) Physics and Chance. Cambridge, UK: Cambridge University Press.

Smoluchowski, M. von (1912) “Experimentell nachweisbare der üblichen Thermodynamik widersprechende Molekularphänomene,” Physik. Z. 13, 1069–1080.

Smoluchowski, M. von (1914) “Gültigkeitsgrenzen des zweiten Hauptsatzes der Wärmtheorie,” in Vorträge übber die Kinetisch Theories der Materie und der Elektrizität (“Limits on the Validity of the Second Law of Thermodynamics,” in Lectures on the Kinetic Theory of Matter and Electricity.), Leipzig, Teubner, pp. 89–121.

\end{document}